# Augmented Networks for Faster Brain Metastases Detection in T1-Weighted Contrast-Enhanced 3D MRI

Engin Dikici[1,*], Xuan V. Nguyen[1], Matthew Bigelow[1], Luciano M. Prevedello[1]

[1] The Ohio State University, College of Medicine, Department of Radiology, Columbus, OH, 43210, USA

[*] Corresponding author: Engin Dikici (engin.dikici@osumc.edu)

## Abstract

Early detection of brain metastases (BM) is one of the determining factors for the successful treatment of patients with cancer; however, the accurate detection of small BM lesions (< 15mm) remains a challenging task. We previously described a framework for the detection of small BM in single-sequence gadolinium-enhanced T1-weighted 3D MRI datasets. It combined classical image processing (IP) with a dedicated convolutional neural network, taking approximately 30 seconds to process each dataset due to computation-intensive IP stages. To overcome the speed limitation, this study aims to reformulate the framework via an augmented pair of CNNs (eliminating the IP) to reduce the processing times while preserving the BM detection performance. Our previous implementation of the BM detection algorithm utilized Laplacian of Gaussians (LoG) for the candidate selection portion of the solution. In this study, we introduce a novel BM candidate detection CNN (cdCNN) to replace this classical IP stage. The network is formulated to have (1) a similar receptive field as the LoG method, and (2) a bias for the detection of BM lesion loci. The proposed CNN is later augmented with a classification CNN to perform the BM detection task. The cdCNN achieved 97.4% BM detection sensitivity when producing 60K candidates per 3D MRI dataset, while the LoG achieved 96.5% detection sensitivity with 73K candidates. The augmented BM detection framework generated on average 9.20 false-positive BM detections per patient for 90% sensitivity, which is comparable with our previous results. However, it processes each 3D data in 1.9 seconds, presenting a 93.5% reduction in the computation time.

## Keywords

Brain metastases; Magnetic resonance imaging; Convolutional neural networks; Computer-aided detection; Scale-space representations

## 1. Introduction

Accurate detection of brain metastases (BM) is critical as the presence of BM implies an advanced and disseminated state of disease. However, if the lesions are detected early when still small, then the disease can be treated with targeted radiotherapy. This may (1) allow for a less invasive and less costly procedure when compared to surgery and (2) lead to fewer adverse neurologic symptoms when compared to whole-brain radiation (Lester et al., 2014). Contrast-Enhanced 3D Magnetic Resonance Imaging (CE-3D-MRI) is commonly used for the detection and monitoring of these lesions. The detection of small BM is often a tedious and time-consuming task for radiologists due to various reasons, including their (1) smaller dimensions, (2) low contrast with surrounding tissues, and (3) visual similarities with vascular structures in some slice angles (Tong et al., 2020).

The automated detection and segmentation of BM via convolutional neural networks (CNNs) in 3D-MRI were investigated in several studies. In (Charron et al., 2018), Charron et al. adapted DeepMedic (Kamnitsas et al., 2017) for the detection of BM in T1-weighted Contrast-Enhanced (T1c) 3D MRI and additional 2D sequences. They reported 7.8 false positives (wrongly detected BM) per study with 93% detection sensitivity for the lesions with an average volume of 2400 mm$^3$. Liu et al. (Liu et al., 2017) introduced En-DeepMedic network for the segmentation of BM in T1c and T2-weighted Fluid-Attenuated Inversion Recovery (FLAIR) images, yielding an average Dice similarity coefficient of 0.67 for tumors with an average volume of 672 mm$^3$. In (Bousabarah et al., 2020), U-Net formulation (Ronneberger et al., 2015) was adapted for the segmentation of smaller BM in T1c, T2-weighted, and T2-weighted FLAIR 3D MRI. The model was evaluated using a dataset with an average tumor size of 1920 mm$^3$ and produced a mean average false positive rate <1 with ~80% BM detection sensitivity. Grøvik et al. (Grovik et al., 2019) employed GoogLeNet (Szegedy et al., 2015) for the segmentation of BM in various sequences (i.e., T1-weighted 3D fast spin echo-CUBE, T1-weighted 3D axial IR-prepped FSPGR, and 3D CUBE FLAIR). They reported 8.3 false positives per study with ~83% detection sensitivity, whereas the dimensional properties of BM in their dataset were not explicitly provided. Cao et al. (Cao et al., 2021) adapted U-Net (Ronneberger et al., 2015) with an asymmetric architecture for the segmentation of BM in T1c 3D MRI, and validated their solution for smaller and larger BM (1-10mm and 11-26 mm in diameter) separately. While the study achieved ~81% detection sensitivity for these smaller BM, it excluded patients with a single BM and tested on a limited group of tumors (i.e., 72 smaller BM in total). More recently, Zhou et al. (Zhou et al., 2020) used deep-learning based single-shot detectors (Liu et al., 2016) for the BM in T1c spoiled gradient-echo 3D MRI; a trained model generates fast inferences (~1s) with a detection sensitivity of 81% and ~6 false positives. In (E Dikici et al., 2020), we proposed a BM detection framework for a single-sequence gadolinium-enhanced T1c 3D MRI. The algorithm generated 9.12 false positives with 90% detection sensitivity for lesions with an average volume of only 160 mm$^3$ (see Fig. 1). However, the process took approximately 30 seconds to complete. A review paper by Cho et al. (Cho et al., 2021) provided comparative analyses of the state-of-art ML-based BM detection studies (including many of the ones mentioned) based on the Checklist for Artificial Intelligence in Medical Imaging (CLAIM) (Mongan et al., 2020) and Quality Assessment of Diagnostic Accuracy Studies (QUADAS-2) criteria (Whiting et al., 2011).

The high performance (i.e., high sensitivity with a relatively low number of false positives) for the detection of particularly tiny BM allows the framework described in (E Dikici et al., 2020) to be a valuable tool, as missing these smaller lesions may potentially compromise the success of treatment planning for the patient. However, the framework's low speed may hinder its integration into modern radiology workflows (Engin Dikici et al., 2020) designed to serve high volumes of computer-aided diagnoses to radiologists. Algorithms with high inference speeds also have the potential to be implemented directly into picture archiving and communication systems (Faggioni et al., 2011), allowing them to bypass the convoluted and inefficient image routing mechanisms. Accordingly, the study reformulates the BM detection algorithm from (E Dikici et al., 2020) using an augmented CNNs approach with the goal of speeding up the algorithm while preserving its accuracy.

This report (1) provides a brief overview of the BM detection framework, (2) describes the Laplacian of Gaussians-based BM candidate selection, constituting a computation bottleneck for the framework, and (3) proposes a candidate detection CNN (cdCNN) to replace it. The results section presents the time consumption and BM detection properties of the proposed solution based on five-fold cross-validation (CV) executed on a cohort of 217 studies. Next, the results are discussed, and the framework is compared with other state-of-art approaches via metrics including the scale of the targeted tumors, detection sensitivity, and processing speed. The report is concluded with a summary of the novelties of the introduced study, system limitations, and future work considerations.

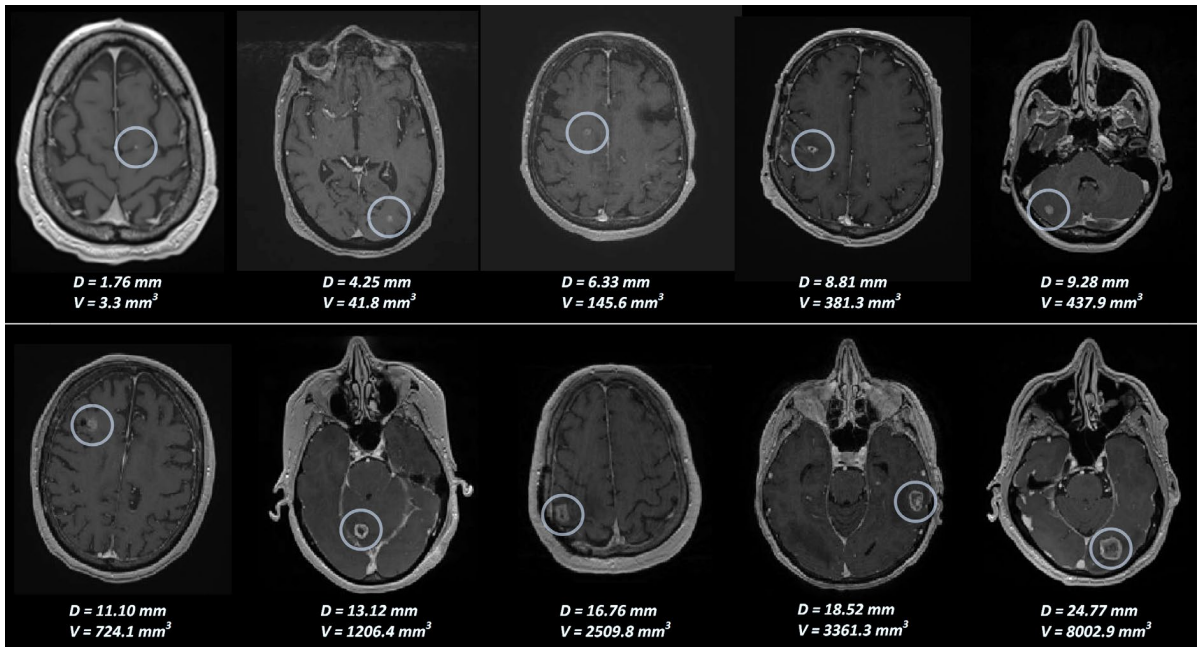

Fig. 1: Example set of BMs with their corresponding diameters (in mm) and volumes (in $mm^3$). Our BM detection framework was trained and validated for BMs with an average volume of $160mm^3$ (with an average diameter of 5.45mm).

## 2. Materials and methods

2.1 BM detection framework

The framework introduced in (E Dikici et al., 2020) consists of two main stages: (1) The BM candidate selection, and (2) the classification of the candidates. The candidate selection stage processes a given volumetric image data to generate a list of 3D coordinates that is likely to contain BM center positions. Accordingly, it adapts the Laplacian of Gaussian (LoG) approach (Lindeberg, 2013) (a widely used blob-shape detection algorithm) with problem specific objectives: (1) Maintaining the detection framework's sensitivity (i.e., the percentage of BMs detected), and (2) limiting the size of generated candidate lists to achieve a higher classification performance. Thus, the optimization process is described as a minimax problem:

$$\arg max_p \big(Sv(LoG(p,V), M)\big), \tag{1}$$

$$\arg min_p (|LoG(p,V)|), \tag{2}$$

where $Sv$ defines the sensitivity of the system based on $M$, the list of correct BM center positions, and $LoG(p,V)$, the candidate points for image data $V$ with LoG parameters of $p$. Accordingly, a sensitivity threshold is integrated as,

$$\arg max_{p, Sv \geq \theta} \big(Sv(LoG(p,V), M)\big), \tag{3}$$

with $\theta$ is the minimal allowed sensitivity, and $p$ is found via grid-search (Wang and Summers, 2012) constrained with Equation-2.

The classification stage of the framework utilizes a dedicated CNN (i.e., CropNet) to determine the probabilities of given candidates to be BM. Due to the under-representation of the BM with respect to other blob-shaped formations; the framework employs random paired sample selections. Random pairs consist of positive and negative samples (i.e., BM centers and candidates that are not BM) that are augmented on-the-fly (Milletari et al., 2016) using random image translations, rotations, flips, gamma-corrections and elastic deformations (Simard et al., 2003) (see Fig. 2). Binary cross-entropy is used as the loss function during the optimization of this classification network.

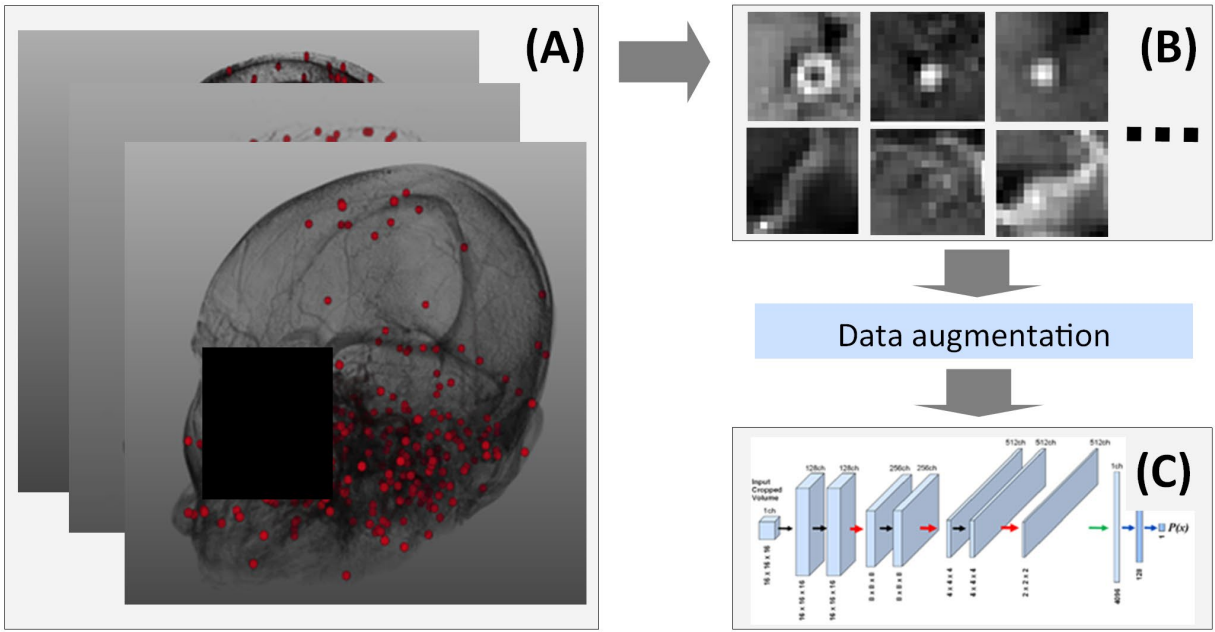

Fig. 2: (A) BM candidates in 3D data are found using the constrained LoG approach. (B) Pairs of positive (i.e., with BM) and negative samples are collected from the candidates. (C) Augmented sample pairs are fed into the classification CNN.

## 2.2. Technical contributions: Candidate detection CNN and network augmentation

The constrained LoG is a CPU-intensive IP technique; it takes 30 seconds to process volumetric data with approximately 16M voxels using recent hardware (see the results section for the specs). This may limit the framework's integration into modern radiology workflows, serving a high volume of calls to deployed computer-aided diagnosis algorithms (Engin Dikici et al., 2020). The computational weight of the approach is due to multiple Gaussian filters applied to a given image. The LoG uses Gaussian filters with a range of standard deviations - SDs ($[\sigma_{min}, \sigma_{max}]$), bounding the detected blob-shaped formations' scale (Lindeberg, 2013). The constrained LoG formulation optimizes this SD range, among other LoG parameters, via Equation-3. The Gaussian kernel radii are determined based on the optimized SDs, where the receptive field ($rf$) is the kernel diameter corresponding to $\sigma_{max}$. For a CNN to replicate a constrained LoG output, it needs to have a matching or greater $rf$: The sensory space of the new approach has comparable dimensions with the constrained LoG.

In this study, we propose a candidate detection CNN (cdCNN) that processes the volumetric data in a fraction of the time required by the constrained LoG. The input for the network is isotropically sampled 3D MRI data with each voxel representing 1mm³. The output is a single channel volumetric data with the same dimensions as the input. The network architecture consists of a stack of dimension-preserving three-channel convolutional blocks; the network's depth $d$ is determined based on the target $rf$:

$$rf = k + (d-1)*(k-1), \tag{4}$$

where $d$ gives the number of sequential convolutional blocks with kernel sizes of $k$ (Araujo et al., 2019).

The input-output pairs for the cdCNN training are prepared as follows: (1) $LoG(p, V)$ is computed for the input $V$ after finding $p$ as described previously, and (2) the corresponding non-smoothed output $Q$ (having the same dimensions as $V$) is set as,

$$Q(x) = \begin{cases} 0 & x \notin LoG(p, V) \\ c \leq 1 & x \in LoG(p, V) \\ 1 & x \in M(V) \end{cases} \tag{5}$$

where (1) $x$ denotes a 3D voxel coordinate, and (2) $c$ is a hyperparameter giving the voxel value for a point that is a candidate point but not a BM center position. Sigmoid activation is used at the output layer to present this [0,1] range output. Please note that $Q$ is a sparse matrix (i.e., ~99.5% of $Q$ is zeros); hence, we chose to use the Dice similarity coefficient as the loss function during the training with a Gaussian smoothed version of the output (i.e., $R = N(Q, \sigma_{smooth})$) to facilitate the model convergence (see Fig. 3).

The conversion of the cdCNN output to a list of 3D points (as the constrained LoG produces) requires the thresholding of the output. The optimal threshold value ($\tau$) is determined by optimizing:

$$\arg max_{\tau, Sv \geq \theta}(Sv(cdCNN(V) > \tau, M)), \tag{6}$$

$$\arg min_\tau(|cdCNN(V) > \tau|), \tag{7}$$

where $cdCNN(V) > \tau$ is used as a shorthand notation for the list of 3D points in $cdCNN$ output that are larger than $\tau$ (all other symbols are adopted from the previous section). More explicitly, Equation-6 maximizes the BM detection sensitivity, whereas Equation-7 minimizes the length of the BM candidates list generated by cdCNN.

TABLE 1: OPTIMAL LOG PARAMETERS, RECEPTIVE FIELD AND NETWORK DEPTH

| CV fold | $\sigma_{min}$ (mm) | $\sigma_{max}$ (mm) | Radius Range (mm)[a] | Receptive Field (mm)[b] | cdCNN depth |
|---|---|---|---|---|---|
| 1 | 1 | 4 | [2,7] | 15 | 7 |
| 2 | 1 | 4 | [2,7] | 15 | 7 |
| 3 | 1 | 5 | [2,9] | 19 | 9 |
| 4 | 1 | 4 | [2,7] | 15 | 7 |
| 5 | 1 | 5 | [2,9] | 19 | 9 |

[a] Gaussian kernel's radius range is derived from the minimal and maximal standard deviations.
[b] Receptive field is the diameter of the Gaussian kernel with $\sigma_{max}$.

A trained cdCNN is used in tandem with the classification network to form a BM detection framework. The study adopts the CropNet network architecture from (E Dikici et al., 2020), where the isotropically sampled input represents a $16mm \times 16mm \times 16mm$ region and produces a binary output giving the BM likelihood probability. It is trained using batches of paired positive and negative samples presenting volumetric regions centered by the correct BM positions, and cdCNN generated candidates that are not BM centers (i.e., away from an actual BM center at least 2mm) respectively. The trained cdCNN and CropNet are deployed in an augmented fashion; the BM candidates generated by cdCNN are fed into CropNet using an iterator for the final BM detections.

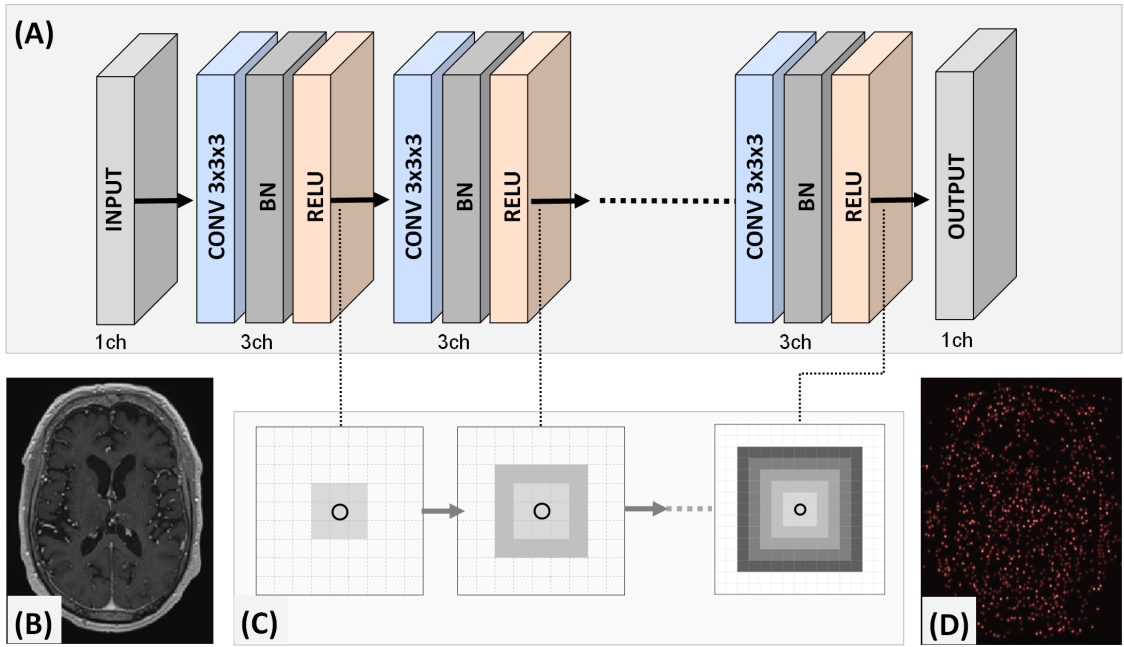

Fig. 3: (A) Architecture of cdCNN, (B) mid axial slice of sample input, (C) receptive field growth for the convolution blocks with kernel size k=3, and (4) mid axial slice of a sample output; showing the centers of blob-shaped formations in the input.

## 3. Results

A total of 217 post-gadolinium T1-weighted 3D MRI exams were collected from 158 patients; 113 patients with a single dataset, 33 patients with 2 datasets (i.e., one follow-up examination), 10 patients with 3 datasets, and 2 patients with 4 datasets. The images were collected from eight scanners, where the acquisition parameters for each are summarized in Appendix-A. Dotarem (i.e., gadoterate meglumine) was used as the contrast agent. The main study selection criterion was the exclusion of datasets involving lesion(s) with a diameter of 15 mm or larger. The data included 932 BM, where the mean BM diameter was 5.45 mm (SD = 2.67 mm), and the mean BM volume was 159.58 mm$^3$ (SD = 275.53 mm$^3$). Appendix-B provides further information on BM included in our data (i.e., the histograms for the BM count per patient, BM diameter, and BM volume).

The evaluation was performed using five-fold cross-validation. The folds were generated patient-based (i.e., each patient was located in either a training or testing group for each fold); hence, the bins included datasets from 31, 31, 32, 32, and 32 patients respectively. For each CV fold, four bins were used for the training and validation, and a single bin was used for the testing.

The constrained LoG algorithm parameters were found for each CV fold via optimizing Equation-3, where the minimal and maximal SDs ($\sigma_{min}$ and $\sigma_{max}$) were searched in the range of $[1, 6]$mm with the step size of 1 mm. The LoG method's Gaussian kernel radius was determined based on the computed SDs ($kernel\ size = \lceil\sqrt{3} \cdot \sigma\rceil$mm as in (Pedregosa et al., 2011)). The cdCNN network consisted of convolution layers with kernel size $k = 3$, which were initialized using Glorot uniform initializer (Glorot and Bengio, 2010). The network depth ($d$) was determined using Equation-4. Table 1 summarizes the LoG SDs, radius range, receptive field, and the corresponding network depth for each CV fold.

Each cdCNN network was trained (1) using Adam algorithm (Kingma and Ba, 2014) with the learning rate of 0.002, (2) for 500 epochs, and (3) using geometrical augmentations on the training data (i.e., image translation, rotation, and flips). The value of hyperparameter $c$ (Equation-5) was determined experimentally: The training was performed for each fold for a set of $c$ values, and the response threshold $\tau$ (Equations 6-7) was adjusted to have ~60K BM candidates on average per study. Fig. 4 shows the BM detection sensitivity of cdCNN with different $c$ values for the training group of each CV fold, where $c = 0.001$ produced the highest sensitivity value of 97.4% (the sensitivity was computed as the percent of actual BM points- $M$ hit by cdCNN generated points with a maximum distance of 1.5mm) and Dice similarity coefficient of 0.84.

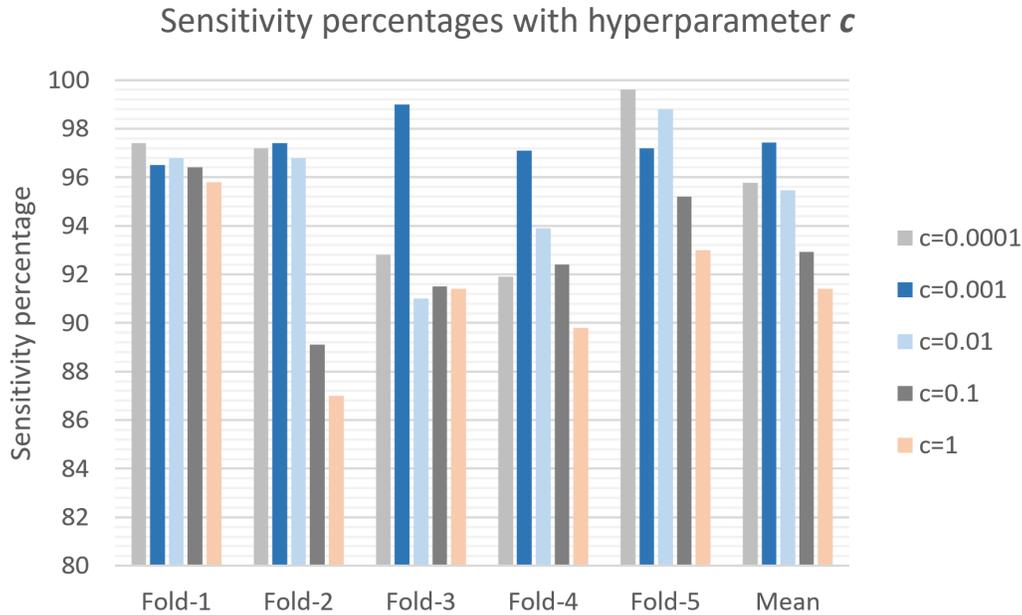

*Fig. 4: The sensitivity percentages of cdCNN computed BM candidates for different c values.*

After the parameter optimizations, cdCNN was augmented into the BM detection framework replacing the constrained LoG stage. The augmented framework was trained (with a similar setup as (E Dikici et al., 2020)) (1) using Adam algorithm with the learning rate of 0.00005, (2) for 20000 batch iterations, where each batch included 130 pairs of positive (BM) and negative (BM candidate yet not a BM) samples. The average number of false positives (i.e. false lesion-detections) per patient (AFP) was computed in connection to the sensitivity of the framework for the testing group of each CV fold. The sensitivity of the framework was adjusted via setting a threshold for the classifier network's response. The augmented framework using cdCNN had the AFP of 9.2 at 90% sensitivity, whereas the constrained LoG used in (E Dikici et al., 2020) had 9.12 at the same sensitivity level. Fig.5 represents the AFP in relation to the detection sensitivity. The computation of BM candidates took 1.1s for each input volume with the cdCNN network, leading to BM detection time of 1.9s for the augmented framework. On the other hand, the constrained LoG-using framework processed each volume in 30.3s, where 29.4s of the time was consumed for the candidate BM detection with the LoG (see Fig.6). Fig.7 shows the BM detection results of the framework with the 90% sensitivity setting for a sample group of unseen data (not used for the training or testing).

The implementations were performed using the Python programming language (v3.6.8). The neural network was created and trained via the Keras library (v2.3.1) with TensorFlow (v1.12.0) backend. The training times for each cdCNN and augmented detection network were ~15.5 and ~2.8 hours, respectively, using a system with NVIDIA Tesla V100 graphics card and Intel Xeon E5-2698 CPU.

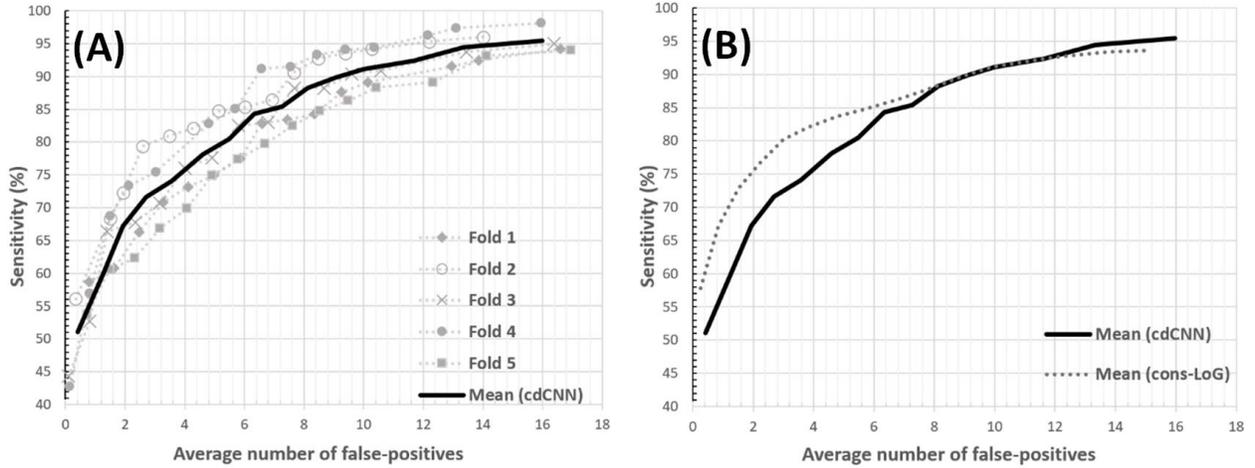

*Fig. 5: (A) Average number of false positives (AFP) per patient (i.e., wrongly detected BM lesions) in relation to the sensitivity is illustrated for the proposed augmented framework. (B) AFP of the augmented framework vs the one using constrained LoG.*

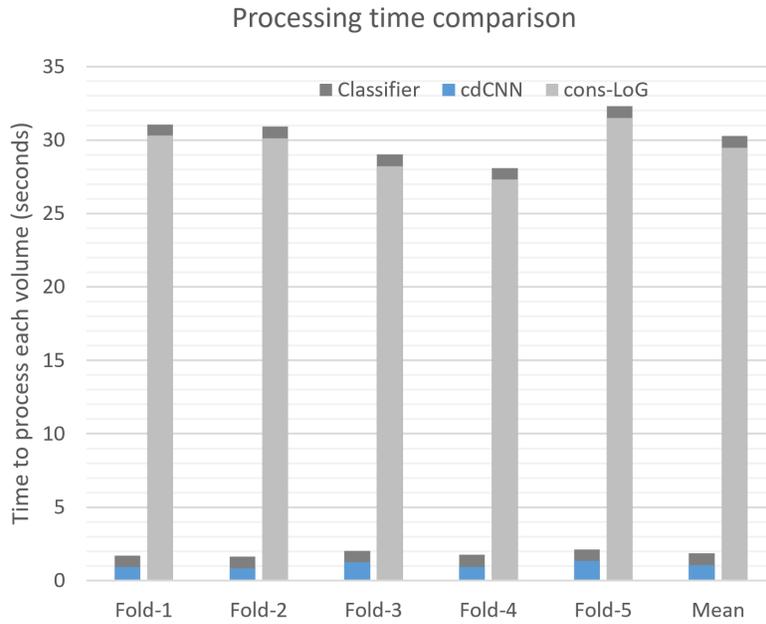

*Fig. 6: Processing time of the augmented framework, using cdCNN and classifier (CropNet), vs the one using constrained LoG and classifier. (The classifier time consumption does not vary significantly as both setups use the same classifier.)*

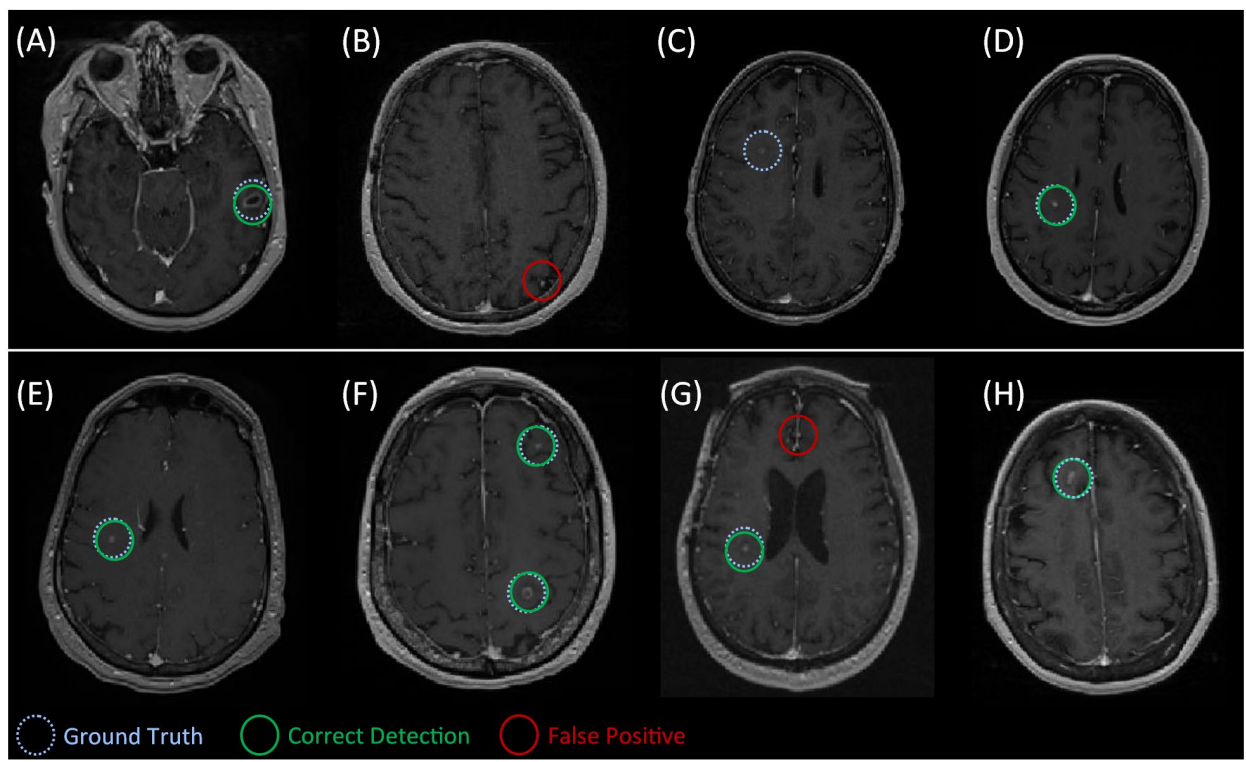

*Fig. 7: Sample set of BM detection results for the proposed framework: In (B) a vessel is incorrectly detected as a BM, in (C) a tumor is missed, and in (G) a vessel adjacent to the cerebral falx incorrectly detected as BM.*

## 4. Discussion

In Fig. 5 (B), we showed the average AFP curves for the proposed augmented framework and the original framework using constrained LoG. The AFP of the augmented network is comparable with the original framework at ≥90% sensitivity, whereas the original framework produces fewer AFPs at lower sensitivity values. As the framework would ideally be deployed with high sensitivity settings (to ensure the detection of a higher percentage of smaller BMs), the behavior of the system at lower sensitivity setups may not cause any practical limitations.

Table 2 provides an overview of the data acquisition types, BM dimensions, validation strategies, detection accuracies, hardware, and 3D MRI data inference times for some of the CNN-based BM-detection/segmentation approaches published over recent years. When compared with the other studies, we utilized a higher performance hardware for the model training and evaluations (excluding (Zhou et al., 2020), which also used a high performance system); hence, the results might have not reflected a fair comparison with other solutions' computational performances. Therefore, we also tested the 3D data processing time of models from this study using the system from (E Dikici et al., 2020), which is equipped with an NVIDIA 1080ti GPU and a 3.5 GHz Intel Core i7-5930K CPU. The proposed framework processed each 3D volume in 3.3 seconds with this configuration.

The cdCNN training output had three classes as shown in Equation-5. If cdCNN was formulated to solely emulate the constrained LoG algorithm, then the equation would host two classes (i.e., member of LoG list, and non-member). However, the cdCNN would underperform as a BM candidate detector, as it would estimate the LoG points without having additional weight for the BM points: In Figure-4 $c = 1$ shows the outcome of this bi-class scenario. Setting $c < 1$ boosts cdCNN for BMs by penalizing the missed BM detections with a higher training loss. Accordingly, cdCNN setups with $c < 1$ achieved higher BM candidate detection sensitivities as displayed in Figure-4.

## 5. Conclusions

The study introduced a BM detection framework for contrast-enhanced 3D MRI datasets, where the framework augmented a novel candidate detection CNN (cdCNN) with a classifier CNN. The study had two main objectives: (1) to maintain the detection performance (i.e., sensitivity and AFP) of a previously defined framework (E Dikici et al., 2020), and (2) increase the processing speed to enable its deployment in demanding radiology workflows and/or radiologist review workstations for on-demand AI-inferences. The results suggest that both of these objectives are achievable with the proposed approach.

A limitation of the study is the requirement of large computational resources: Training of a single cdCNN took ~15.5h using a highly advanced GPU (NVIDIA Tesla V100) as reported. Alternative CNN formulations may mitigate hardware requirements and speed up the training process. Employing dilated convolutions (Wei et al., 2018) would lead to a remarkably shallower, and therefore faster training, cdCNN; having a dilation rate of two (in all three dimensions) in all convolution layers would reduce the depth of a cdCNN with a target receptive field of 19 from 9 to 5. The reduction in depth may introduce a reduced network expressiveness (Eldan and Shamir, 2016) and generalizability (Liu et al., 2020), yet the concept is open to further investigation in our problem domain. Additionally, the usage of residual connections (He et al., 2016) may speed up the training convergence speed of the network. These and further modifications and corresponding ablation studies may be a topic for a future study.

TABLE 2: OVERVIEW OF BM SEGMENTATION/DETECTION APPROACHES FOR 3D MRI USING CNNS

| Study | Acquisition | BM volume (mm³) | Validation type | Sensitivity | AFP | GPU | Inference Time |
|---|---|---|---|---|---|---|---|
| Charron et al. (Charron et al., 2018) | Multi seq.[a] | Mean: 2400 Median: 500 | Fixed train/test | 93 | 7.8 | NVIDIA 1080 | 20 min |
| Liu et al. (Liu et al., 2017) | Multi seq.[b] | Mean: 672 | 5-fold CV | NA | NA | NVIDIA Quadro M2000M | 2 min |
| Bousabarah et al. (Bousabarah et al., 2020) | Multi seq.[c] | Mean: 1920 Median: 470 | Fixed train/test | 77-82 | < 1 | NVIDIA 2080ti | 4.5 min |
| Grøvik et al. (Grovik et al., 2019) | Multi seq.[d] | NA | Fixed train/test | 83 | 8.3 | NVIDIA 1080ti | 1 min |
| Cao et al. (Cao et al., 2021) | T1cMRI | NA [g] | Fixed train/test | 81-100 [j] | NA | 4 x NVIDIA 2080ti | NA |
| Zhou et al. (Zhou et al., 2020) | T1c MRI [e] | NA [h] | Fixed train/test | 81 | 6 | NVIDIA V100 | 1 sec |
| Dikici et al. (E Dikici et al., 2020) | T1c MRI | Mean: 160 Median: 50 | 5-fold CV | 90 | 9.12 | NVIDIA V100 NVIDIA 1080ti | 30.3 sec 31.5 sec |
| **This study** | **T1c MRI [f]** | **Mean: 160 Median: 50** | **5-fold CV** | **90** | **9.20** | **NVIDIA V100 NVIDIA 1080ti** | **1.9 sec 3.3 sec** |

[a] *T1-weighted 3D MRI with gadolinium injection, T2-weighted 2D FLAIR, and T1-weighted 2D MRI sequences.*
[b] *T1c and T2-weighted FLAIR sequences.*
[c] *T1c, T2-weighted, and T2-weighted FLAIR sequences.*
[d] *Pre- and post-gadolinium CUBE, post-gadolinium T1-weighted 3D axial IR-prepped FSPGR (BRAVO), and 3D CUBE FLAIR sequences.*
[e] *3D T1-weighted contrast enhanced spoiled gradient-echo MRI sequence.*
[f] *The same dataset was used for training and validation in this study and [13].*
[g] *Tumor volumes were not reported. The testing was done for 72 smaller tumors (1-10mm in diameter) and 17 larger tumors (11-26mm in diameter) separately.*
[h] *Tumor volumes were not reported. The average size of tumors in the study were 10mm ± (standard deviation: 8mm).*
[j] *Sensitivity was 81 percent for smaller tumors and 100 percent for larger ones.*

The optimization of the network (e.g., parametrizing the depth, filter counts, filter sizes, etc.) could benefit from neural architecture search algorithms (Elsken et al., 2019). However, these methods commonly require the training of the network with many possible variations; hence, they might not be feasible considering the long training time of a cdCNN even with the aforementioned modifications. A recent study by Mellor et al. (Mellor et al., 2020) proposed to analyze activation patterns of networks without any network training for fast optimizations, producing promising results. The application of such architecture optimization schemes in our problem domain might be investigated in a later project.

The segmentation of BM is important for the dimensional quantification of the tumors and their monitoring over time (Gaonkar et al., 2015). It is also a valuable tool for medical interventions such as stereotactic radiosurgery, which requires accurate segmentation/contouring of the tumors (Liu et al., 2017). The proposed framework is currently limited to the detection aspect of the problem, but it could be adapted for the segmentation of BM in a future study. One way to achieve this might be replacing the classifier with a segmentation network such as (Milletari et al., 2016; Ronneberger et al., 2015), trained with an applicable loss type (e.g., Dice loss). Alternatively, activation maps (Zhou et al., 2016) of the classifier network may also be utilized for a coarse segmentation of the tumors as described in (Natekar et al., 2020).

The main technical contribution of the project, the replacement of computationally heavy constrained LoG algorithm with a dedicated CNN, can be utilized in various other applications employing classical IP for pre-processing and/or problem-specific feature extraction purposes (e.g., (Junior et al., 2018; Mao et al., 2021)). Accordingly, we will continue to elaborate on the augmentation of CNNs (and NNs) with different objectives for the fast and accurate processing of medical images in the future.

## Disclosures

No conflicts of interests, financial or otherwise, are declared by the authors.

# Appendix-A

TABLE 3: SCANNER PARAMETERS

| Scanner | MF [a] (T) | TR [b] range (ms) | TE [c] range (ms) | Slice thickness range (mm) | Pixel size [d] range (mm) | Imaging frequency range (MHz) | Flip angle range (degrees) | Exam # |
|---|---|---|---|---|---|---|---|---|
| Siemens Aera [e] | 1.5 | [9.3, 9,7] | [4.4, 4.7] | 1.0 | [0.78, 0.97] | 63.6 | 20 | 54 |
| Siemens Avanto [e] | 1.5 | [9.7, 10] | [4.2, 4.8] | [0.9, 1.0] | [0.43, 0.86] | 63.6 | [15, 20] | 17 |
| Siemens Espree [e] | 1.5 | 10 | [4.5, 4.7] | 1.0 | [0.78, 1.0] | 63.6 | 20 | 26 |
| Siemens Skyra [e] | 3.0 | [6.2, 6.5] | 2.46 | [0.8, 0.9] | [0.65, 0.78] | 123.2 | [10.5, 12] | 34 |
| Siemens TrioTrim [e] | 3.0 | 6.5 | 2.45 | [0.8, 0.9] | [0.65, 0.73] | 123.2 | 10.5 | 4 |
| Siemens Verio [e] | 3.0 | [6.5, 9.0] | [2.4, 4.9] | [0.8, 0.9] | [0.65, 0.78] | 123.2 | 10.5 | 28 |
| GE Optima MR450w [f] | 1.5 | 10.3 | 4.2 | 1.0 | 0.49 | 63.9 | 20 | 26 |
| GE Signa HDxt [f] | 1.5 | [9.2, 10.3] | 4.2 | 1.0 | [0.49, 0.98] | 63.9 | 20 | 28 |

[a] *Magnetic field strength,* [b] *repetition time,* [c] *echo time,* [d] *pixel size is same in x and y directions.*
[e] *Siemens Healthcare, Erlangen, Germany.*
[f] *GE Healthcare, Milwaukee, Wisconsin, USA.*

# Appendix-B

The database included 932 BMs where,

- The mean number of BMs per patient is 4.29 (σ = 5.52),
- The median number of BMs per patient is 2,
- The mean BM diameter is 5.45 mm (σ = 2.67 mm),
- The median BM diameter is 4.57 mm,
- The mean BM volume is 159.58 mm3 (σ = 275.53 mm3),
- The median BM volume is 50.40 mm3.

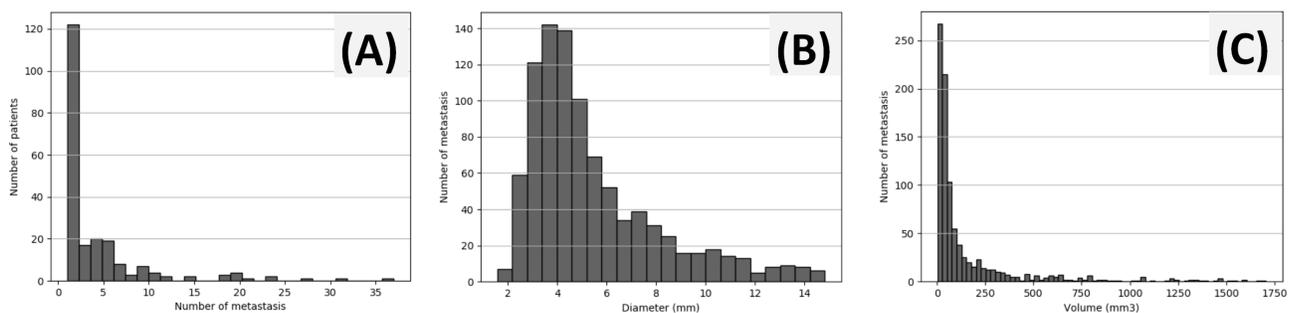

*Fig. 6 Histograms for the BM (A) count per patient, (B) diameter and (C) volume.*